# Insights into the Non-exponential Behaviour of the Dielectric Debye-like Relaxation in Monoalcohols

S. Arrese-Igor[1], A. Alegría[1,2], A. Arbe[1] , and J. Colmenero[1, 2, 3]

[1] *Centro de Física de Materiales (CSIC-UPV/EHU), Materials Physics Center (MPC), Paseo Manuel Lardizabal 5, 20018 San Sebastián, Spain*

[2] *Departamento de Física de Materiales UPV/EHU, Apartado 1072, 20080 San Sebastián, Spain*

[3] *Donostia International Physics Center DIPC, Paseo Manuel Lardizabal 4, 20018 San Sebastián, Spain*

**Abstract:** More than 100 years after Debye proposed his model for the dielectric relaxation of monoalcohols (MA), some fundamental questions about their dynamics and its relation with the supramolecular structures created by hydrogen bonding remain not fully understood. The recent detection of the dynamics of hydrogen-bonded aggregates by techniques other than dielectric spectroscopy is leading to novel insights and opening new questions. In particular it was recently reported that the shear response of some MA present three relaxation components, at variance with their apparent bimodal dielectric response. We show by mixing an archetype MA with LiCl that both shear and dielectric measurements are consistent with the presence of three separated contributions. Results are discussed in the more general context of MAs with non-exponential slow dielectric relaxations. In particular, we propose the recently reported third process as the origin of the unusual broadening of the Debye-like dielectric relaxation of some MAs.

Hydrogen bonding has significant impact on the thermodynamics, the structuring and the dynamics of many important fluids with biological or technical applications. From very structured fluids like colloids or surfactants to hydrogen bonded molecular liquids, the emergence of interaction-mediated characteristic mesoscale structures are a paramount ingredient on their particular properties. In this context, deep understanding of the relationship between structure and dynamics is strongly desirable. Monoalcohols (MA) are a typical example of hydrogen bonded liquids with rich dynamics, which have been extensively studied by multiple techniques [1]. Particularly remarkable is their dielectric relaxation, which in addition to the structural relaxation shows an intriguing very intense slower relaxation reported more than 100 years ago, commonly referred to as 'Debye' relaxation. Generally, this relaxation is very close to an exponential, a Debye function in the frequency domain, and hence its name. There exist, however, MAs with non-exponential or time distributed 'Debye' relaxations, and therefore, we will use the term Debye-like to refer to the intense slower dielectric relaxation observed in these systems. The typical morphology of the mesoscale structures leading to a pre-peak in the static structure factor of MAs would be chain-like and ring-like structures as identified from computer simulations, with the majority of molecules forming chain-like aggregates [2–6]. The orientational correlation alignment of dipoles forming chains would be at the origin of the intense Debye-like dielectric relaxation, that according to the transient chain model[7] would basically relax through successive loss and/or gain of segments at the end of chains. Although it is broadly accepted that the Debye-like dielectric relaxation is related to supramolecular structures or aggregates resulting from hydrogen bond interactions, there is no evident explanation why this relaxation is broad for some MAs. It seems clear, however, that the degree of non-exponentiality of the Debye-like relaxation is correlated with its strength, so that the broader the relaxation the less intense it is [8,9].

For long time the observation of the Debye-like process by other techniques than dielectric spectroscopy remained elusive, but slow dynamics additional to that of the structural relaxation have already been reported by a number of techniques [10–15] and most revealingly by shear modulus measurements [16–18]. Moreover, very recently and aided by complex viscosity $\eta^*(\omega)$ representation of shear data, authors have reported for the first time the presence of an additional process, intermediate between those previously indentified [19]. The timescale of this additional feature observed in the shear viscosity response is consistent with the timescale of the rotational dynamics of hydroxyl groups as seen by NMR [7,20]. As a consequence, this intermediate process was interpreted as the mechanical imprint of the breaking and forming of individual hydrogen bonds in the framework of the transient chain model. As illustrated in figure 1, the multi- modal shear response of MAs in general shows the following three separate processes: (i) a slow relaxation leading to the recovery of pure viscous flow; (ii) the mentioned additional process at intermediate frequencies; and (iii) a high frequency maximum representative of the structural relaxation involved in the glass transition. Comparison of these results with those obtained by dielectric spectroscopy becomes very interesting. On the one hand, the high frequency peak in $\eta''(\omega)$ matches that of the dielectric $\alpha$ relaxation, confirming its assignment to the structural relaxation. On the other hand, the characteristic time where the liquid recovers pure viscous flow (crossover to $\eta''(\omega) \propto \omega$) correlates well with that of the dielectric Debye-like relaxation in many MAs [17,19,21,22], supporting the idea that the dielectric Debye-like relaxation reflects the complete relaxation of hydrogen-bonded aggregates. The intermediate process observed in the shear response, however, seems not to have a clear dielectric analogue. The reorientation of individual OH groups attaching and detaching the aggregate could in principle lead to dielectric signal. Interestingly, long be- fore we reported the presence of the intermediate process in the shear response of MAs, we had noticed that when digging into great detail the description of the dielectric relaxation of 2-ethyl-1-hexanol (2E1H) in terms of two processes provided insufficient intensity at the valley between the two relaxations, opening the possibility of an additional dielectric relaxation process in between [23]. Further exploring these ideas, the purpose of this work was to modify hydrogen bond interactions and structures with several objectives: gain more knowledge on the origin and nature of the mentioned intermediate process; search for possible traces and indications of intermediate process on the dielectric response of MAs; and find an explanation for the non-exponential behavior of some MAs.

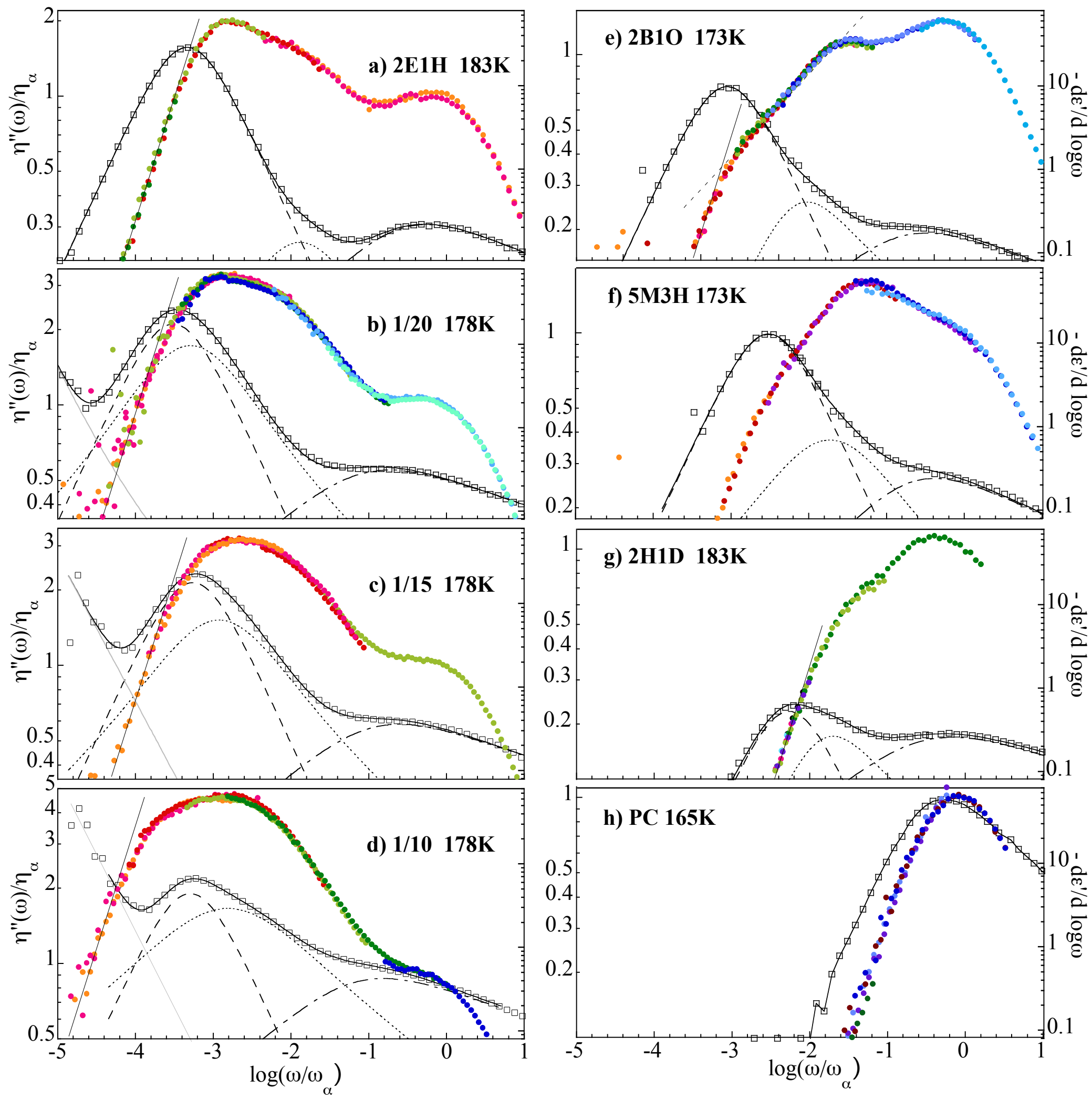

FIG. 1: Shear master curves (left axis -circles-) and isothermal dielectric measurements (right axis -squares-) at the indicated reference temperature for pure 2B1O, 5M3H, 2H1D, 2E1H MAs, LiCl/2E1H mixtures and PC. Solid lines in shear data represent the terminal behavior $\eta''(\omega) \propto \omega$. Solid, dashed, dotted and dotted-dashed lines in DS data represent total, Debye, intermediate, and a relaxation fit components, respectively -see SI-.

The mixing of MAs with other polar and non-polar liquids and even with other MAs has been a usual approach to modify hydrogen bonds and better understand their dielectric response[24–31]. In most cases relaxation in the mixture is broadened with respect to the pure MA and Debye-like process amplitude decreases. One mayor problem when dealing with such mixtures is that it be- comes difficult to distinguish between several broadening scenarios, namely: (i) intrinsic broadening of the Debye-like relaxation of the MA component; (ii) general broadening of relaxations due to mixing effects, a common phenomenology in low molecular weight mixtures and polymer blends [32–35]; and (iii) apparent broadening due overlapping of contributions from different components. In order to avoid the extra level of complexity introduced by the dynamics of a second liquid we have studied in this work the evolution of the mechanic and dielectric response of 2E1H upon LiCl salt addition. Among different MA the choice of 2E1H results convenient due to the relatively large timescale separation between the structural and the Debye-like relaxation in this MA and its clear multimodal shear response [17,19]. Solutions of LiCl in 2E1H were prepared at 1/30, 1/20, 1/15, and 1/10 LiCl/2E1H molar ratios (between 1% and 3% weight percentage). The resulting solutions were characterized by means of: infrared spectroscopy in the near (NIR) and medium (MIR) ranges; small angle X-ray scattering (SAXS); differential scanning calorimetry (DSC); dielectric spectroscopy (DS); and oscillatory shear response - see Supporting Information (SI) for details-.

We found first evidences of the changes produced by LiCl addition on the hydrogen bonded mesoscopic structure of 2E1H in FTIR and SAXS measurements ( see SI ). On the one hand, LiCl addition modifies OH-group vibrations confirming that ions do interact with hydroxyl hydrogens. The red shift of the stretch band in the MIR indicates further weakening of the OH covalent bond due to dipolar interaction with ions. In the NIR range mixtures show a decrease of both free (~ 7092 $cm^{-1}$) and strongly associated (~ 6280 $cm^{-1}$) OH bonds [36,37], consistent with an scenario where ions reduce the number of hydroxyl groups available, hampering the formation of strongly bonded associates. Structural characterization by SAXS on the other hand, shows that the mesoscopic pre-peak characteristic of hydrogen bonded aggregates moves to lower scattering vector (Q) values - equivalently, larger characteristic length scales - as LiCl content increases, whereas main peak characteristic of the intemolecular correlations between CH groups [3] remains invariable. The position of the mesoscopic pre-peak in MA (mainly corresponding to O-O correlations) is known to move to lower Q the larger the number of carbons in the main chain[3]. Here being the alkyl chain length the same, a shift to lower Q values is compatible with a decrease in the density of aggregates. An increase of the X-ray scattering contrast due to the presence of ions interacting with end-chain OH-units is also compatible with the observed increase in the relative intensity of the pre-peak.

Panels a) to d) of figure 1 show the imaginary part of the shear viscosity $\eta''(\omega)$ master curves for pure 2E1H and LiCl-containing 2E1H at different concentrations. The y-scale in figure 1 covers only one decade in order to underline different maxima while still being sensible to data $\propto \omega^x$ power laws. Non-simple, multimodal, relaxation is observed for all the samples [19]. The higher frequency maximum should be representative of the glass transition while the rest of the features presumably originate from mesoscale structures and/or interactions between molecules. As commented, three different processes or components can be identified for pure 2E1H. The addition of LiCl decreases the relative intensity of the slowest and more prominent component in 2E1H, the more, the higher the LiCl content. Regarding dielectric spectroscopy results, mixing mainly results in broadening of the slowest relaxation peak together with a gradual decrease of its intensity (see figures 1, 2 and SI for additional representation of dielectric data). The overall dielectric response of 2E1H/LiCl mixtures thus resembles the not frequently reported non-exponential behavior of the Debye-like relaxation observed for some MA, like 2- butyl-1-octanol (2B1O), 2-hexyl-1-decanol (2H1D), or 5- methyl-3-heptanol (5M3H) [8,9], illustrated in figure 1 as well together with the simple liquid propylene carbonate (PC) for comparison.

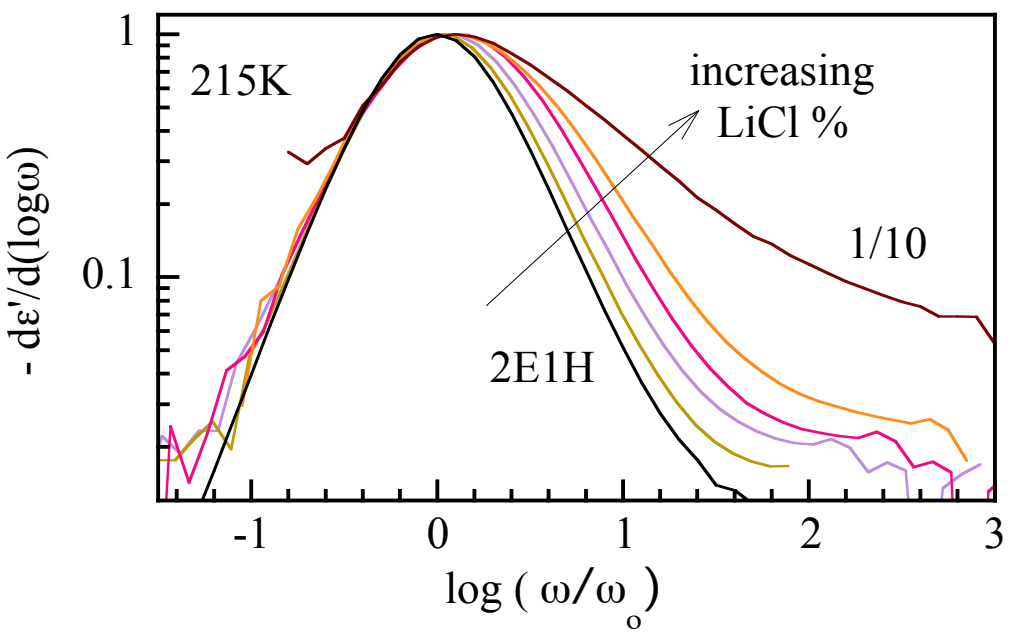

FIG. 2: Maximum normalized derivative of the real part of the permittivity for different salt concentrations at 215K.

The evolution of the dielectric signal of 2E1H upon LiCl addition confirms that small ions are able to modify the 2E1H aggregates responsible of the prominent slow dielectric relaxation of the alcohol, and therefore act as 'structure breakers'. Concomitant with the changes in the slow dielectric relaxation and the increase of the characteristic length scale of the mesoscopic structures ob- served by SAXS, the shear response of 2E1H mixtures also gets modified upon LiCl addition. In this case the shoulder at intermediate frequencies in $\eta''(\omega)$ (which we will refer to as 'intermediate process') seems to gain importance relative to the other two. The effect of ion addition tested by different techniques altogether, leave little room against the interpretation of the Debye-like and shear terminal relaxations as due to the relaxation of hydrogen bonded aggregates.

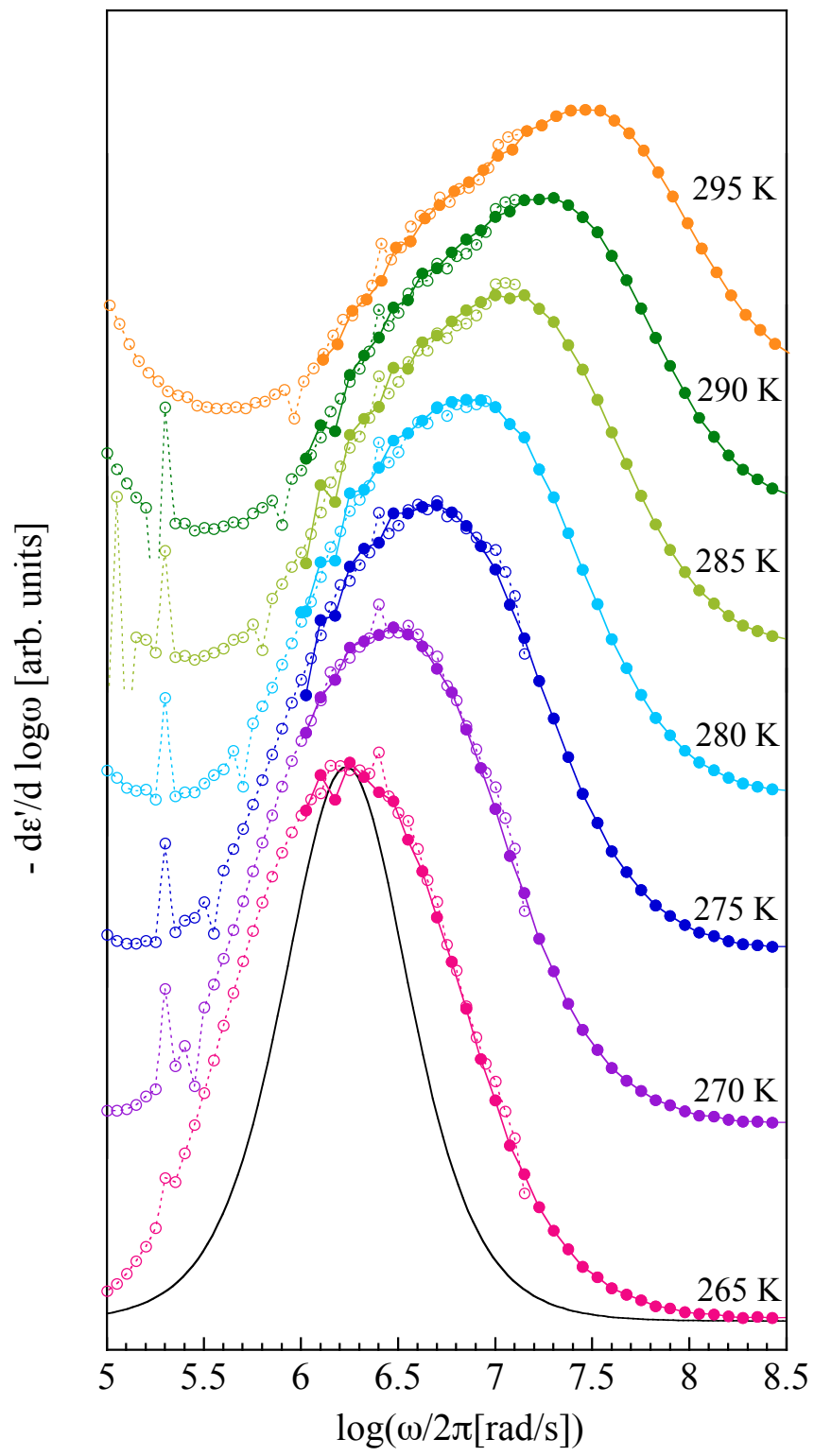

FIG. 3: Temperature evolution of the slowest peak of the derivative of the real part of the dielectric permittivity from 265K to 295K every 5K for the 1/15 molar ratio sample. Curves corresponding to different temperatures were shifted for clarity. Solid line represents an exponential (Debye) relaxation.

In addition to test the effect of the addition of a 'structure breaker' on the dynamics of 2E1H, in this work we also aimed at modifying the relative contribution of the slowest and intermediate processes to the response obtained by different techniques. In particular, one of the objectives was to make more evident the very subtle sign of the intermediate process in the dielectric relaxation [9,19,23] previously proposed by us. For pure 2E1H the width of the slowest dielectric relaxation slightly increases as temperature increases, or otherwise, as intensity decreases. This tendency seems to be quite universal [8,9] and could be interpreted as an indication of the presence of an underlying process whose contribution appears more clear when the intensity of the slow Debye relaxation decreases due to less or smaller hydrogen bonded aggregates. In principle, if LiCl breaks non-covalent interactions, similar phenomenology could be expected here, that is, a decrease of the intensity together with a broadening of the overall signal due to the presence of an underlying process. In this regard, LiCl/2E1H mixtures present lower intensity slow relaxations together with a progressive broadening of the signal as salt concentration increases. As shown in figure 2, upon LiCl addition there is a considerable broadening on the high-frequency flank of the relaxation, i.e. where the contribution of the intermediate process is expected, whereas the low-frequency tail remains more or less the same. Moreover, for a constant salt concentration the effect of increasing temperature is that observed in pure 2E1H, i.e. further broadening of the slow dielectric signal. This is again consistent with the above mentioned picture where the signature of an underlying process (an increase of the relaxation width in this case) appears more clear as the intensity of the slow Debye relaxation decreases. As can be seen in figure 3, the change in the shape of the slow dielectric signal of 1/15 molar sample as temperature increases is dramatic. At the lowest temperatures the signal is already broader than a pure exponential relaxation and as temperature increases it continues broadening in a non-symmetric way, so that for the highest temperatures two separate processes could be envisaged. At very high temperatures the intensity of the slowest Debye-like relaxation decreases to such extent that the underlying additional process (intermediate between Debye-like and α relaxations) seems to dominate the dielectric response, so that the asymmetric broadening faces now towards the low-frequency flank.

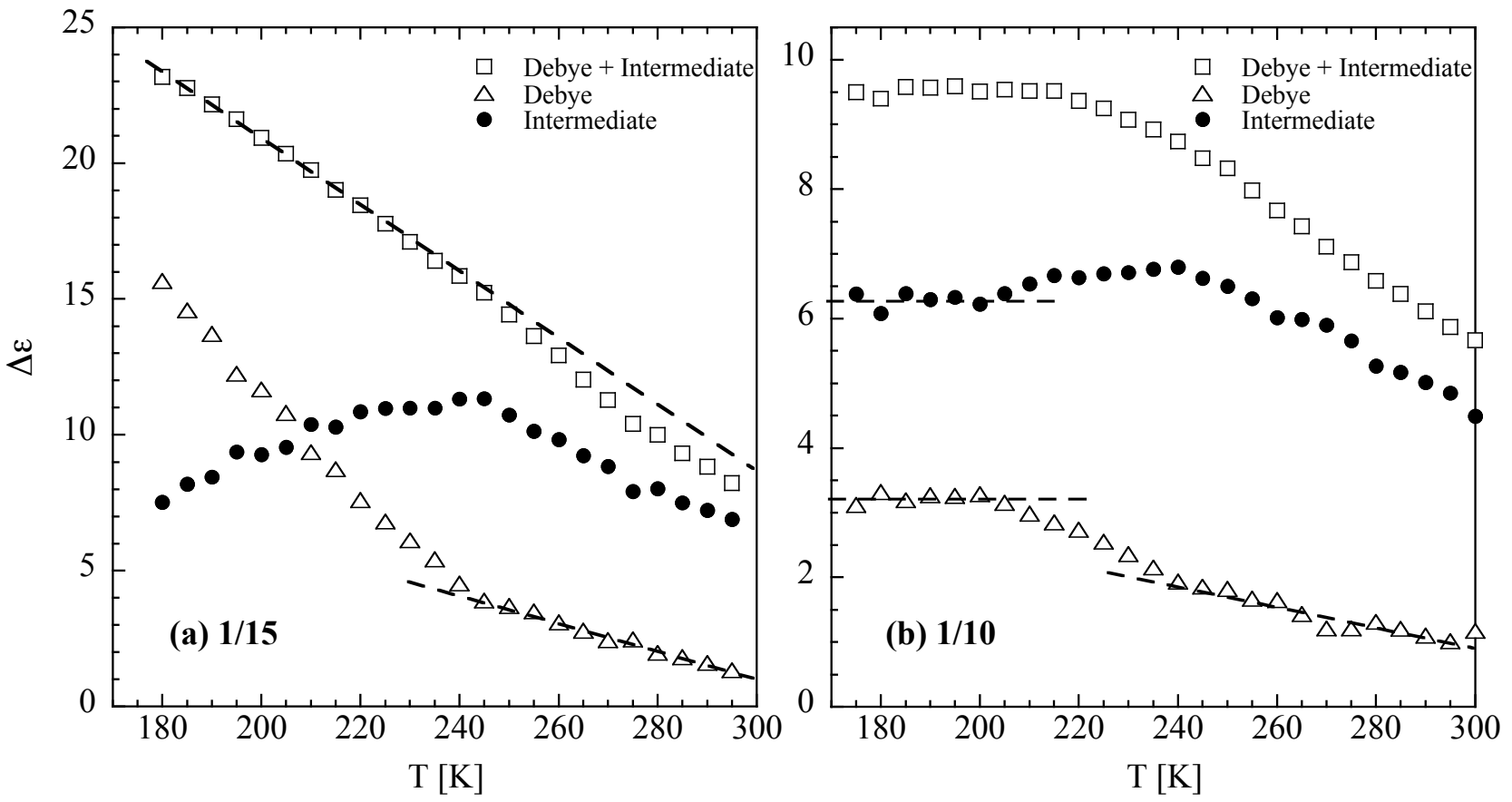

FIG. 4: Temperature evolution of the dielectric strength for 1/15 (panel a) and 1/10 (panel b) molar ratio samples. Dashed lines are guides for the eye highlighting changes of trend.

The presence of an intermediate dielectric process can rationalize the effect of salt concentration and T- evolution on the shape of the slow dielectric feature in 2E1H/LiCl mixtures and is consistent with the modification of the shear response upon salt addition. As commented before, the timescale for the intermediate shear process in pure 2E1H agrees well with the timescale ob- served by NMR for the hydroxyl group [7,20] roughly reflecting the lifetime of a molecule within a transient chain of alcohol molecules. The presence of 'structure breaker' ions would in principle increase the relative number of 'terminal' monomers, increasing the contribution of the breaking of individual hydrogen bonds to the different experimental observables. In order to dig into this idea we made an approach to describe dielectric data in terms of a sum of three processes (see SI for details). Obtained dielectric strengths for 1/15 and 1/10 samples are shown in figure 4. As in 2E1H [38], the T-dependence of the strength of both the Debye and the intermediate processes in 1/15 sample show a change of trend at about 240-250K (see Figure 4a). This change has also been observed for other physical quantities as absorbance of first overtone stretches of OH and chemical shift of the hydroxyl proton measured by NMR, and for other MAs as well[38]. Despite the strength of both processes reflect changes at ~240-250K, their temperature behavior is distinct. The strength of the Debye process continuously decreases with temperature, while that of the intermediate process shows a maximum about 240-250K. The occurrence of a maximum in the strength of the inter- mediate process associated to the breaking and forming of individual hydrogen bonds on the aggregates would be the result of two competing effects. On the one hand, the decrease of the average chain-like cluster size with increasing temperature[38] and salt concentration increases the relative number of 'terminal monomers', increasing the contribution of the intermediate process to the different experimental observables. This effect, however, cannot be extended forever. At some point, the gradual extinction of the H-bonded aggregates by itself precludes the occurrence of the intermediate process, which requires cooperation or interaction between OH moieties, as the independent dielectric relaxation of OH groups should reflect the structural α-relaxation. Consistently, the strength of the intermediate process in the 1/10 ratio sample (panel -b- of Figure 4) remains constant as long as that of the Debye process does, i.e. as long as there is not a significant decrease on the number or size of H- bonded aggregates. Destruction of H-bonded aggregates (decrease of Debye relaxation strength) above ~ 200 K produces an increase in the attaching/detaching of individual hydrogen bonds until the strength of the inter- mediate process reaches a maximum simultaneous to the change of trend in the decrease of the Debye relaxation strength at ~ 240 K.

Although the presented experimental work deals with mixtures of LiCl/2E1H, the proposed interpretation is general and should in principle be extensible to other MAs (and even other H-bonded liquids). Particularly interesting is the generalization for those MAs with modest strength and non-exponential relaxations like 2B1O, 5M3H, and 2H1D shown in figure 1. As in 2E1H, the shear response of these MAs shows three processes [19]. The frequency of the highest and lowest features in the shear response match the maxima of the α and the Debye- like dielectric relaxations respectively. In the proposed scenario, the 'extra' broadening of the Debye-like relaxation at the high frequency wing would be nothing else than the dielectric counterpart of the intermediate process seen in the shear response. Once again, the three processes inferred from shear measurements are qualitatively consistent with the dielectric response and its interpretation as due to three separated contributions. More- over, this rationalization provides a natural explanation for the correlation between the extent of broadness and the strength of the Debye-like relaxation, as the presence of an underlying process becomes more evident the lower the intensity of the slowest pure Debye component.

In conclusion, we have proved that LiCl ionic salt mixed with 2E1H acts as a 'structure breaker', modifying both the dielectric relaxation and the mechanical response of the MA and shifting the diffraction structural pre-peak characteristic of mesoscopic structures. All the experimental observations agree with an scenario where the presence of ions limits the formation of strongly associated hydrogen bonded aggregates. The impact of LiCl addition on the three relaxation components identified in the shear response is uneven, and the evolution of the dielectric signal with temperature and LiCl concentration is consistent with the presence of three separated contributions in the dielectric response as well. Generalization of the proposed scenario provides a natural explanation for the unusual spectral broadening of some MAs, which would just reflect the dielectric signature of the intermediate process observed in the shear response. In a more general context, it is reasonable to think that similar dynamic processes can also take place in other hydrogen bonded systems, where structural relaxation, individual hydrogen bond forming and breaking and complete relaxation of the hydrogen bonded aggregates should to a greater or lesser extent dictate their dynamic response depending on the bonding lifetime and architecture of the structures formed.

## Acknowledgments

We acknowledge financial support from the projects: PGC2018-094548-B-I00 by the Spanish Ministry "Ministerio de Ciencia, Innovación y Universidades" (MICINN- Spain) and FEDER-UE; and IT-1175-19 by the Basque Government.

## References

[1] R. Böhmer, C. Gainaru and R. Richert. Structure and dynamics of monohydroxy alcohols - Milestones towards their microscopic understanding, 100 years after Debye. *Physics Reports* **2014**, 545, 125-195.

[2] J. L. MacCallum, and D. P. Tieleman. Structures of Neat and Hydrated 1-Octanol from Computer Simulations. *J. Am. Chem. Soc.* **2002**, 124, 15085.

[3] M. Tomšič , A. Jamnik, G. Fritz-Popovski, O. Glatter, and L. Vlček. Structural properties of pure simple alcohols from ethanol, propanol, butanol, pentanol, to hexanol: comparing Monte Carlo simulations with experimental SAXS data. *J. Phys. Chem. B* **2007**, 111, 1738.

[4] P. Sillrén, J. Bielecki, J. Mattsson, L.Börjesson and A. Matic. A statistical model of hydrogen bond networks in liquid alcohols. *J. Chem. Phys.* **2012**, 136, 094514.

[5] R. Ludwig. The Structure of Liquid Methanol. *Chem. Phys. Chem.* **2005**, 6, 1369.

[6] J. Lehtola, M. Hakala and K. Hämäläinen. Structure of liquid linear alcohols. *J. Phys. Chem. B* **2010**, 114, 6426- 6436.

[7] C. Gainaru, R. Meier, S. Schildmann, C. Lederle, W. Hiller, E. A. Rössler, and R. Böhmer. Nuclear-Magnetic Resonance measurements reveal the origin of the Debye process in monohydroxy alcohols. *Phys. Rev. Lett.* **2010**, 105, 258303.

[8] Y. Gao, W. Tu, Z. Chen, Y. Tian, R. Liu, and L.M. Wang. Dielectric relaxation of long-chain glass-forming monohydroxy alcohols. *J. Chem. Phys.* **2013**, 139, 164504

[9] S. Arrese-Igor, A. Alegría, and J. Colmenero. On the non-exponentiality of the dielectric Debye-like relaxation of monoalcohols. *J. Chem. Phys.,* **2017**, 146, 114502.

[10] T. Yamaguchi, M. Saito, K. Yoshida, T. Yamaguchi, Y. Yoda, and M. Seto. Structural relaxation and viscoelasticity of a higher alcohol with mesoscopic structure. *Phys. Chem. Lett.* **2018**, 9, 298-301.

[11] T. Yamaguchi, A. Faraone, M. Nagao. Collective Mesoscale Dynamics of Liquid 1-Dodecanol Studied by Neutron Spin-Echo Spectroscopy with Isotopic Substitution and Molecular Dynamics Simulation. *J. Phys. Chem. B*, **2019**, 123, 239-246.

[12] J. Gabriel, F. Pabst, A. Helbling, T. Böhmer, and T. Blochowicz. Nature of the Debye-Process in Monohydroxy Alcohols: 5-Methyl-2-Hexanol Investigated by Depolarized Light Scattering and Dielectric Spectroscopy. *Phys. Rev. Lett.,* **2018**, 121, 035501.

[13] J. Gabriel, F. Pabst, and T. Blochowicz. Debye-Process and β-relaxation in 1-Propanol probed by Dielectric Spectroscopy and Depolarized Dynamics Light Scattering. *J. Phys. Chem. B*, **2017**, 121, 8847-8853.

[14] U. Kaatze. Dielectric and structural relaxation in water and some monohydric alcohols. *J. Chem. Phys.* **2015**, 147, 024502.

[15] T. Cosby, A. Holt, P. J. Griffin, Y. Wang, and J. Sangoro. Proton Transport in Imidazoles: Unraveling the Role of Supramolecular Structure. *Phys. Chem. Lett.* **2015**, 6, 3961-3965.

[16] Y. Wang, P. J. Griffin, A. Holt, F. Fan, and A. P. Sokolov. Observation of the slow, Debye-like relaxation in hydrogen-bonded liquids by dynamic light scattering. *J. Chem. Phys.* **2014**, 140, 104510.

[17] C. Gainaru, R. Figuli, T. Hecksher, B. Jakobsen, J. C. Dyre, M. Wilhelm, and R. Böhmer. Shear modulus investigations of monohydroxy alcohols: evidence for a short-chain-polymer rheological response. *Phys. Rev. Lett.*, **2014**, 112, 098301.

[18] T. Hecksher and B. Jakobsen. Supramolecular structures in monohydroxy alcohols: Insights from shear-mechanical studies of a systematic series of octanol structural isomers. *J. Chem. Phys.* **2014**, 141, 101104.

[19] S. Arrese-Igor, A.Alegría, and J. Colmenero. Multimodal character of shear viscosity response in hydrogen bonded liquids. *Phys. Chem. Chem. Phys.* **2018**, 20, 27758-27765.

[20] S. Shildmann, A. Reiser, R. Gainaru, and R. Böhmer. Nuclear magnetic resonance and dielectric noise study of spectral densities and correlation functions in the glass forming monoalcohol 2-ethyl-1-hexanol. *J. Chem. Phys.* **2011**, 135, 174511.

[21] S. P. Bierwirth, C. Gainaru, and R. Böhmer. Communication: Correlation of terminal relaxation rate and viscosity enhancement in supramolecular small-molecule liquids. *J. Chem. Phys.* **2018**, 148, 221102.

[22] S. P. Bierwirth, G. Honorio, C. Gainaru, and R. Böhmer. Linear and nonlinear shear studies reveal supramolecular responses in supercooled monohydroxy alcohols with faint dielectric signatures. *J. Chem. Phys.* **2019**, 150, 104501.

[23] S. Arrese-Igor, A. Alegría, and J. Colmenero. Dielectric relaxation of 2ethyl-1hexanol around the glass transition by thermally stimulated depolarization currents. *J. Chem. Phys.*, **2015,** 142, 214504.

[24] S. S. N. Murthy and Madhusudan Tyagi. Experimental study of the high frequency relaxation process in monohydroxy alcohols. *J. Phys. Chem.* **2002**, 117, 3837-3847.

[25] L-M. Wang, S. Shahriari, and R. Richert. Diluent effects on the Debye-type dielectric relaxation in viscous monohydroxy alcohols. *J. Phys. Chem. B* **2005**, 109, 23255-23262.

[26] L. Hennous, A. R. A. Hamid, R. Lefort, D. Morineau, P. Malfreyt and A. Ghoufi. Crossover in structure and dynamics of a primary alcohol induced by hydrogen-bonds dilution. *J. Chem. Phys.* **2014**, 141, 204503.

[27] T. El Goresy and R. Böhmer. Diluting the hydrogen bonds in viscous solutions of n-butanol with n-bromobutane: A dielectric study. *J. Chem. Phys.* **2012**, 128, 154520.

[28] M. Preuß, C. Gainaru, T. Hecksher, S. Bauer, J. C. Dyre, R. Richert and R. Böhmer. Experimental studies of Debye-like process and structural relaxation in mixtures of 2-ethyl-1-hexanol and 2-ethyl-1-hexyl bromide. *J. Chem. Phys.* **2012**, 137, 144502.

[29] S. Bauer, H. Wittkamp, S. Schildmann, M. Frey, W. Hiller, T. Hecksher, N. B. Olsen, C. Gainaru, and R. Böhmer. Broadband dynamics in neat 4-mthyl-3-heptanol and in mixtures with 2-ethyl-1-hexanol. *J. Chem. Phys.* **2013**, 139, 134503.

[30] S. P. Bierwirth, T. Büning, C. Gainaru, C. Sternemann, M. Tolan, and R. Böhmer. Supramolecular x-ray signature of susceptibility amplification in hydrogen-bonded liquids. *Phys. Rev. E.* **2014**, 90, 052807.

[31] L. P. Singh, A. Raihane, C. Alba-Simionesco, and R. Richert. Dopant effects on 2-ethyl-1hexanol: A dual- channel impedance spectroscopy and neutron scattering study. J. Chem. Phys. 2015, 142, 014501.

[32] J. Colmenero and A. Arbe. Segmental dynamics in miscible polymer blends: recent results and open questions. *Soft Matter* **2007**, 3, 1474-1485.

[33] J. Swenson, and S. Cerveny. Dynamics of Deeply Super- cooled Interfacial Water. *J. Phys.: Condens. Matter* **2015**, 27, 033102.

[34] D. Cangialosi, A. Alegría, and J. Colmenero. Dielectric relaxation of polychlorinated biphenyl/toluene mixtures: Component dynamics. *J. Chem. Phys.* **2008**, 128, 224508.

[35] G. A. Schwartz, M. Paluch, A. Alegría, and J. Colmenero. High pressure dynamics of polymer/plasticizer mixtures. *J. Chem. Phys.* **2008**, 131, 044906.

[36] l. Stordrange, A. A. Christy, O. M. Kvalheim, H. Shen, Y. Liang. Study of the self-association of alcohols by near- infrared spectroscopy and multivariate 2d techniques. *J. Phys. Chem. A* **2002**, 106, 8543-8553.

[37] M. A. Czarnecki and K. Orzechowski. Effect of temperature and concentration on self-association of octan-3-ol studied by vibrational spectroscopy and dielectric measurements *J. Phys. Chem. A* **2003**, 107, 1119-1126.

[38] S. Bauer, K. Burlafinger, C. Gainaru, P. Lunkenheimer, W. Hiller, A. Loidl and R. Böhmer. Debye relaxation and 250K anomaly in glass forming monohydroxy alcohols. *J. Chem. Phys.* **2013**, 138, 094505.

-----------------------------------------------------------------------------------------

**SUPPORTING INFORMATION: Insights into the Non-exponential Behavior of the Dielectric Debye-like Relaxation in Monoalcohols; Experimental details and results**

S. Arrese-Igor[1], A. Alegría[1,2], A. Arbe[1] , and J. Colmenero[1, 2, 3]
[1] *Centro de Física de Materiales (CSIC-UPV/EHU), Materials Physics Center (MPC), Paseo Manuel Lardizabal 5, 20018 San Sebastián, Spain*
[2] *Departamento de Física de Materiales UPV/EHU, Apartado 1072, 20080 San Sebastián, Spain*
[3] *Donostia International Physics Center DIPC, Paseo Manuel Lardizabal 4, 20018 San Sebastián, Spain*

## A. VIBRATIONAL SPECTROSCOPY CHARACTERIZATION

Vibrational spectra of samples were recorded by means of a Jasco 6300 spectrometer in transmission configuration for the near infrared range (8000-4000 $cm^{-1}$) and in Attenuated Total Reflectance (ATR) configuration for the medium infrared range (4000-700 $cm^{-1}$) at room temperature. Figure S1 shows that LiCl addition modifies OH group vibrations: stretching band in the medium infrared (MIR) range and stretch overtones and combination bands in the near infrared (NIR), confirming that ions do interact with hydroxyl hydrogens. The red shift of the stretch band in the MIR indicates further weakening or destabilization of the OH covalent bond due to dipolar interaction with ions. In the NIR range and according to previous assignment [1,2], mixtures show a decrease o both free (~ 7092 $cm^{-1}$) and strongly associated (~ 6280 $cm^{-1}$) OH bonds. These results are consistent with an scenario where ions act as 'structure breakers' reducing the number of hydroxyl groups available and hampering the formation of strongly bonded associates.

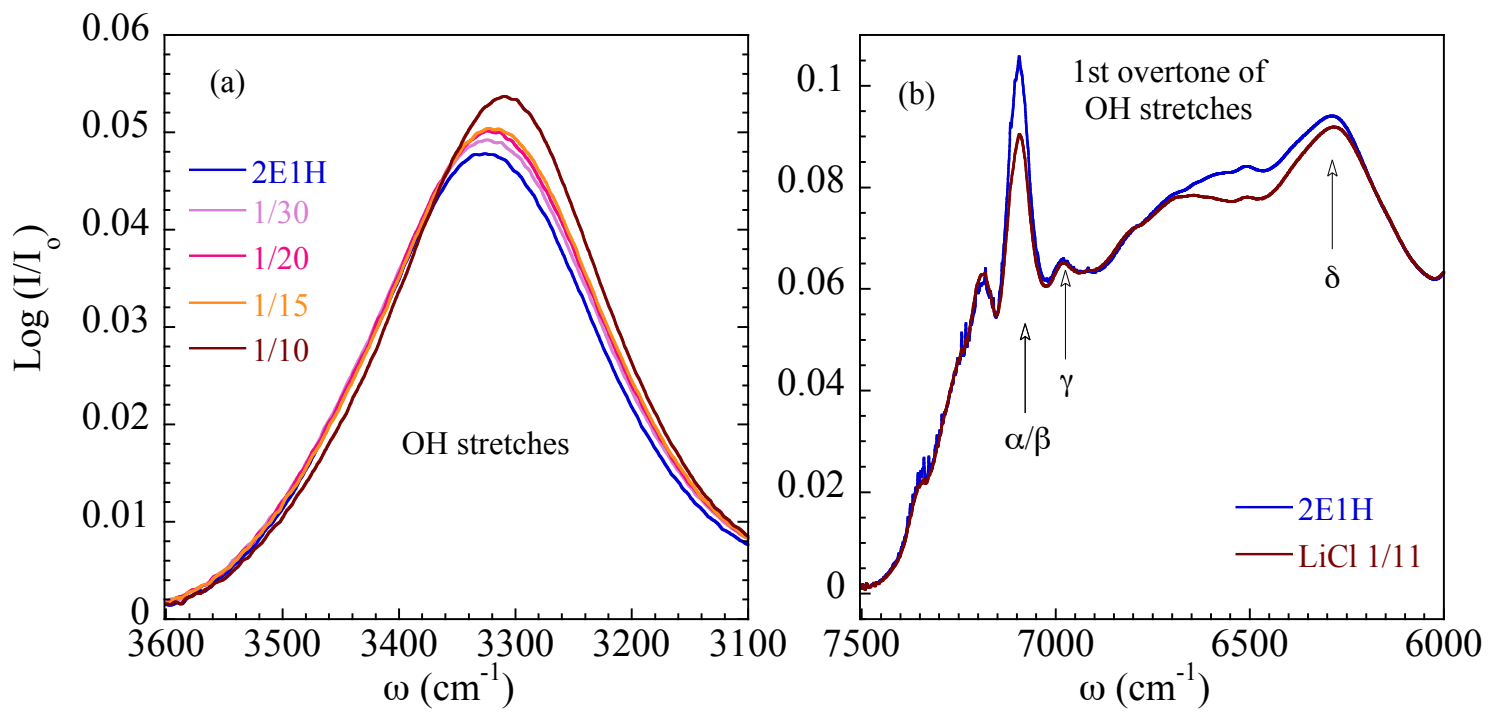

FIG. S1: Infrared absorption spectra obtained for pure 2E1H and mixtures with different LiCl content. Panel (a): MIR range; Panel (b): NIR range.

## B. STRUCTURAL CHARACTERIZATION

Small angle X-ray scattering (SAXS) diffraction experiments were carried out by means of a Rigaku 3-pinhole PSAXS-L equipment operating at 45 kV and 0.88 mA. CuKα transition photons of wavelength $\lambda$ = 1.54 Å were produced by a MicroMax-002+ X-Ray Generator System composed by a microfocus sealed tube source module and an integrated X-Ray generator unit. Flight path and sample chamber in the equipment were under vacuum. A two-dimensional multiwire X-Ray Detector (Gabriel design, 2D-200X) detected the scattered X-Rays. This gas-filled proportional type detector offers a 200 mm diameter active area with ca. 200 micron resolution. The azimuthally averaged scattered intensities were obtained as a function of wavevector Q, $Q = 4\pi/\lambda \cdot \sin\theta$, where $2\theta$ is the scattering angle. Boron-rich capillaries of 2mm thickness were filled with sample and placed in transmission geometry with a sample to detector distance of 23 cm. With this configuration the Q-range covered was $0.07 \leq Q \leq 1.7$ Å$^{-1}$. All the experiments were carried out at room temperature and results are shown in figure S2.

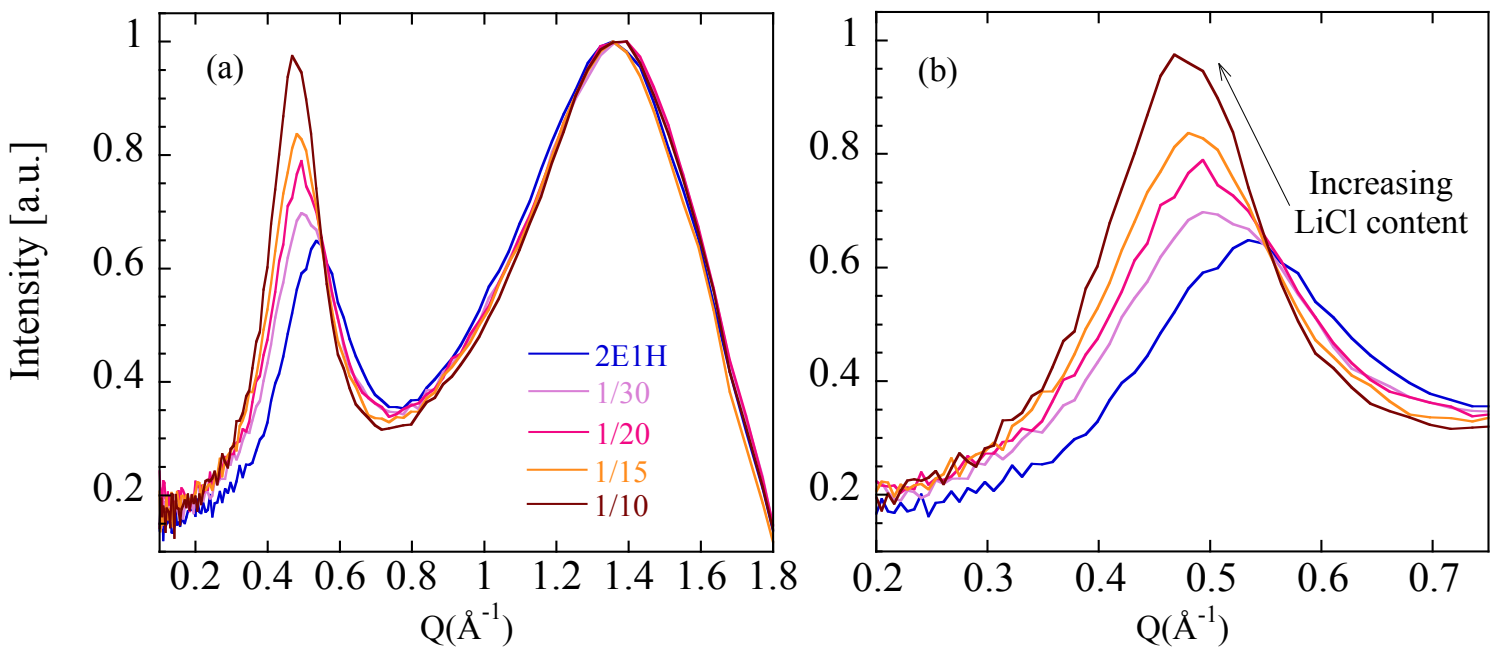

FIG. S2 SAXS diffraction curves obtained for pure 2E1H and mixtures with different LiCl content at ambient temperature. Scattered intensity for different samples was normalized to the maximum of the main peak at ~ 1.4 Å$^{-1}$

## C. DYNAMIC CHARACTERIZATION

### C1. Differential Scanning Calorimetry

Calorimetric glass transition of the samples was determined from the reversible part of the heat flow measured by Temperature Modulated Differential Scanning Calorimetry (Q2000 TAInstruments) at a heating rate of 3K/min and 0.3K modulated amplitude during a 45s period. Figure S3 shows the derivative of the reversible heat flow for pure 2E1H and LiCl/2E1H mixtures at 1/30, 1/20, 1/15, and 1/10 molar ratios. Salt addition produces a slight increase of the glass transition temperature ($T_g$). The mentioned increase is linear with salt concentration and amounts a maximum of 5-6 degrees for the highest molar ratio mixture 1/10.

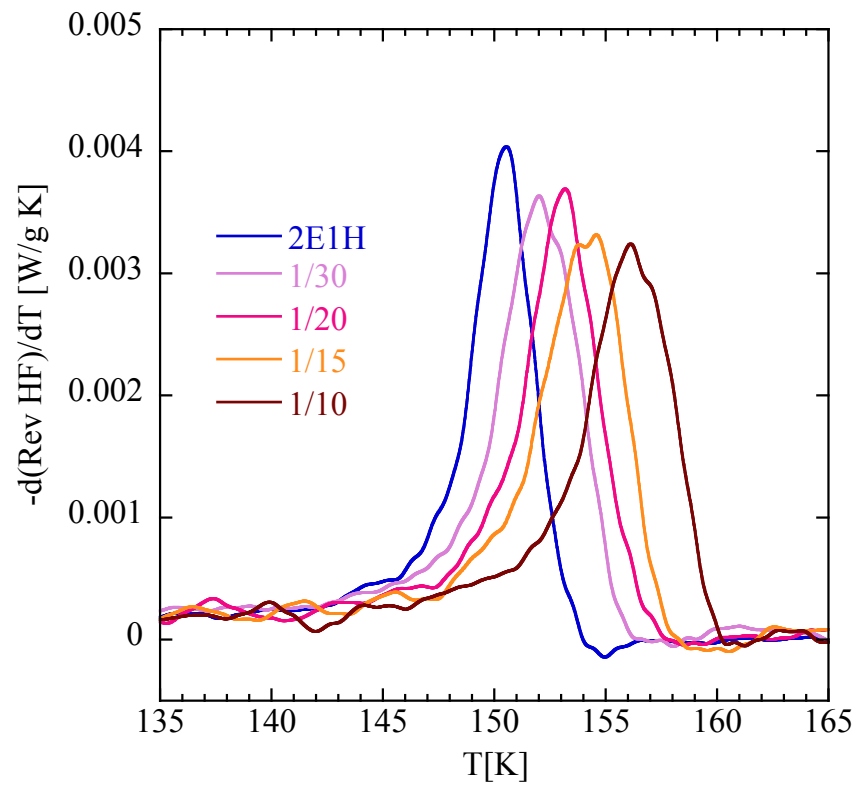

FIG. S3: Derivative of the reversible heat flow for pure 2E1H and LiCl/2E1H mixtures at 1/30 (purple), 1/20 (pink), 1/15 (orange), and 1/10 (brown) molar ratios.

### C2. Shear Response

The shear response was measured by means of a TAInstruments ARES rheometer with a separate motor transducer using invar plates (8 mm diameter) in parallel geometry. Each sample was analyzed over a range of temperatures from about 2 degrees below the calorimetric glass transition to a temperature high enough for the terminal behavior to be observed, generally around 15 - 25 degrees above $T_g$. At the lowest temperature, dynamic strain sweep tests at 200 rad/sec were performed to determine the linear regime in which the storage (G') and loss (G") moduli were constant for each system and a strain in this linear regime was chosen for subsequent measurements which provided torque in the desired range. Dynamic frequency sweeps at a constant strain were run at angular frequencies between 0.1 and 100 rad/sec. Shear data were analyzed paying attention to both the complex modulus ($G^* = G' + iG''$) and complex viscosity ($\eta^* = \eta' - i\eta''$) representations, which are related as $G^* = i\omega\eta^*$. Although the information contained in both magnitudes is the same, the $\eta^*$ representation can aid in resolving the presence of different processes by eye as they may appear highlighted in the form of maxima (or more pronounced power law changes or 'shoulders'). Master curves were first constructed al- lowing horizontal shift of shear modulus data at different temperatures. From these modulus master curves, viscosity master curves were later calculated by means of $\eta^* = G^*/i\omega$. Shear viscosity master curves in figure 1 of the manuscript were normalized in the x-axis, to the structural relaxation timescale taken as the frequency where $G'(\omega_\alpha) = G''(\omega_\alpha)$; and in the y-axis, to $\eta_\alpha = G(\omega_\alpha)/\omega_\alpha$.

### C3. Dielectric Response

Following the procedure applied in our previous works [3,4] the information from broadband dielectric spectroscopy (BDS) experiments was mainly obtained from the analysis of the log-frequency derivative of the real part of the dielectric permittivity $\varepsilon'(\omega)$. In this way: (i) we minimize the uncertainty in the determination of the parameters characterizing the $\alpha$ relaxation and its possible influence on the width of resolved Debye-like process; and (ii) eliminate dc-conductivity contribution from the relaxation peaks. Figure S4 shows that by representing the derivative of the real part of the permittivity slow relaxation peaks are nicely resolved and well separated from electrode polarization effects. The spectra in figures S4 and S5 show the decrease of the relaxation strength of the slowest peak upon salt addition, a decrease that is relatively larger at low temperatures (or larger characteristic time scales) than at high temperatures (or shorter characteristic timescales). Complementary to figure 3 of the manuscript, figure S6 shows the change in the shape and broadness of the slowest dielectric peak for 1/15 mixture in the whole frequency range covered. Figure S7 summarizes the evolution of the dielectric strength of the samples parametrized as ($\varepsilon_0 - \varepsilon_\infty$) where $\varepsilon_\infty$ was taken as the high frequency value of $\varepsilon'(\omega)$ measured at the lowest temperature (125 K), and $\varepsilon_0$ as the low frequency plateau value of $\varepsilon'(\omega)$ at each temperature. The so obtained dielectric strength does not discriminate among the different contributions to the relaxation of the polarization but it is model independent, unlike the dielectric strengths ($\Delta\varepsilon$) obtained from fitting procedures. Data were represented as a function of the inverse of the frequency of the maximum of the Debye-like relaxation ($\omega_{max} = 1/\tau_{max} = 2\pi f_{max}$) to eliminate the effect of the weak slowing down of the dynamics with salt concentration.

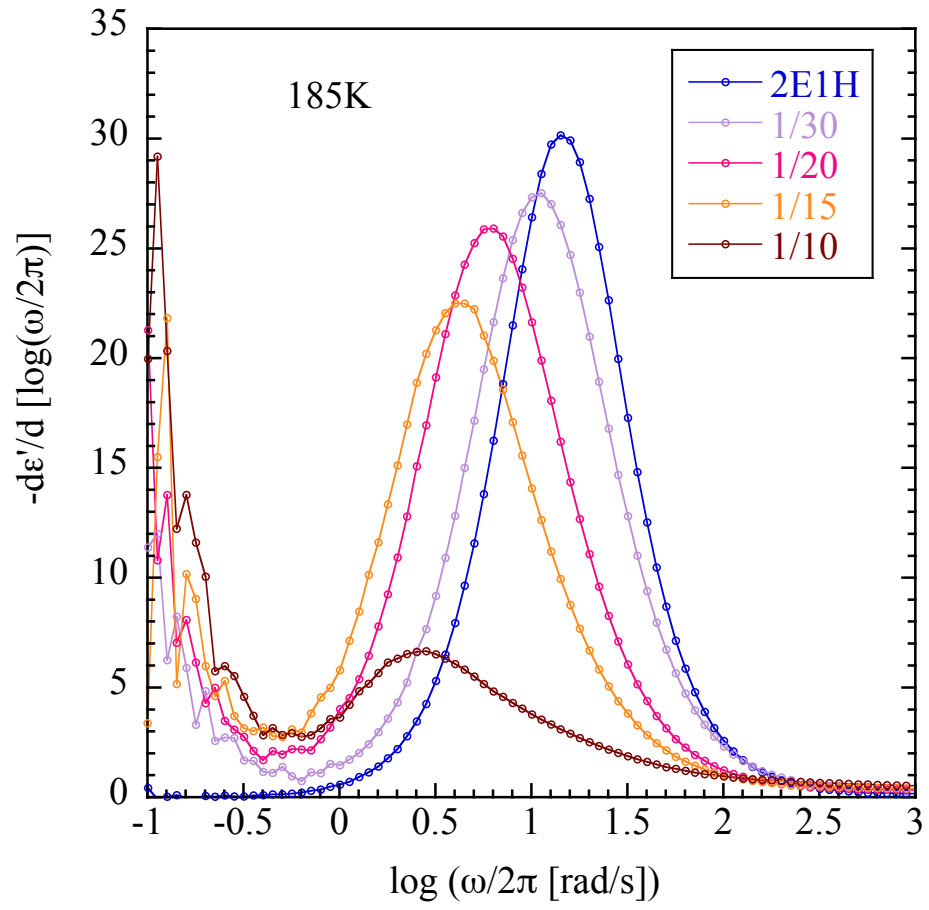

FIG. S4: Derivative of the real part of the dielectric permittivity for different salt concentrations at 185K.

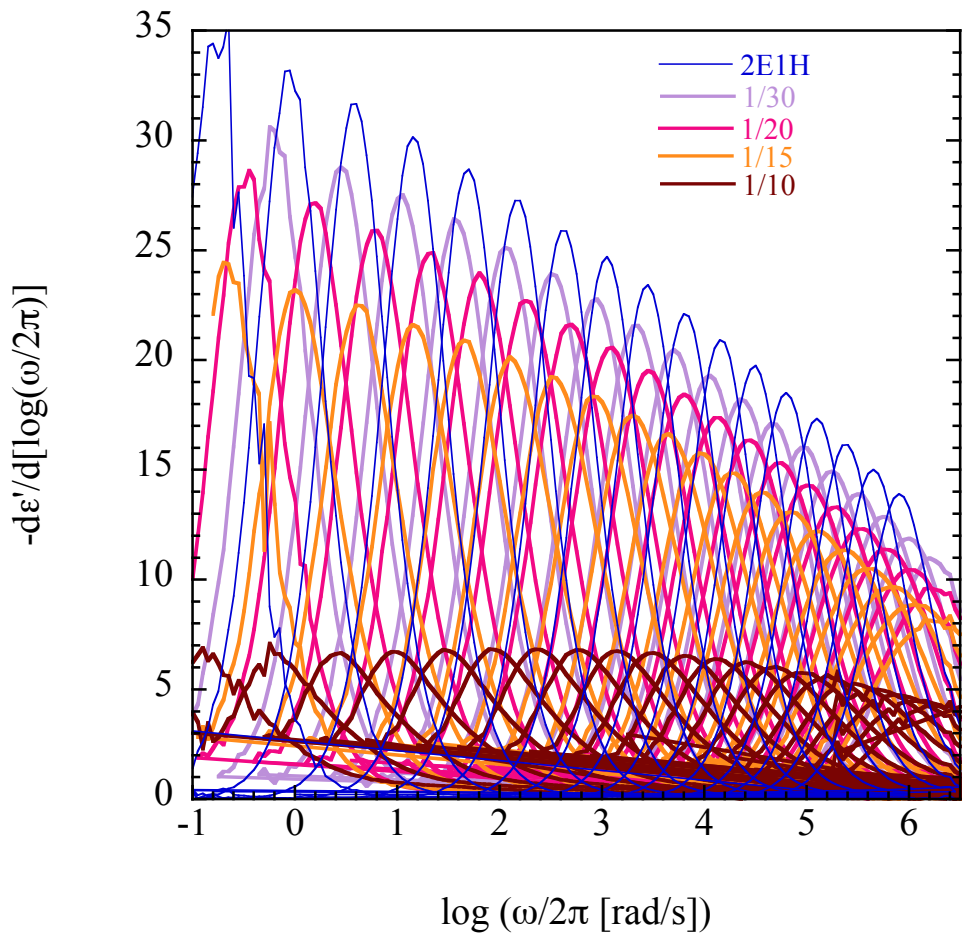

FIG. S5: Temperature evolution of the derivative of the real part of the permittivity for different salt concentrations.

Finally, figure S8 shows the effect of LiCl concentration on the broadening of the slowest relaxation of the mixtures. For this purpose we chose to represent Full Width at Half Maximum (FWHM) value of the derivative of the real part of the permittivity, which is a model-free parameter. As commented in the body of the manuscript, the width of the slowest dielectric relaxation for pure 2E1H slightly increases as temperature increases, or otherwise, as intensity decreases [3,4]. Note that this tendency for 2E1H is not well captured when the width is quantified in terms of FWHM, as the broadening is low and it occurs at the high-frequency tail of the relaxation. Data in figure S8 were again represented as a function of the inverse of the frequency of the maximum of the Debye- like relaxation ($\omega_{max} = 1/\tau_{max} = 2\pi f_{max}$) to eliminate the effect of the weak slowing down of the dynamics with salt concentration.

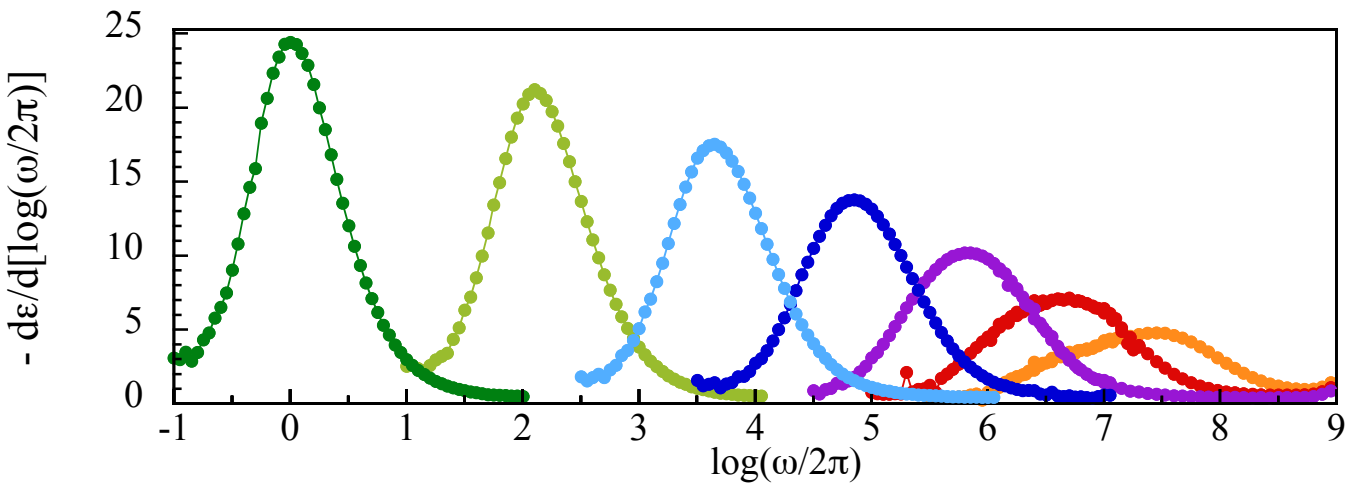

FIG. S6: Temperature evolution of the derivative of the real part of the dielectric permittivity from 175K to 295k every 20K for the 2E1H/LiCl 1/15 molar ratio mixture.

Quantitative description of dielectric data was done as follows. As in our previous works, when analyzing the derivative of the real part of the dielectric permittivity we did not assume the approximation $\varepsilon'' \sim d\varepsilon'/d\log\omega$, but the complex permittivity was modeled by a sum of Havriliak-Negami (HN) functions [5] as usual,

$$\varepsilon^*(\omega) = \varepsilon_\infty + \sum \Delta\varepsilon_j \; / \; [1 + (i\omega\tau_j)^{a_j}]^{b_j} , \qquad (1)$$

where each process j is characterized by its relaxation strength $\Delta\varepsilon = \varepsilon_o - \varepsilon_\infty$, its relaxation time $\tau$, and the shape parameters $a$ and $b$ ($0 < a, b \leq 1$) determining the broadening and asymmetry of the loss curve respectively. The derivative of the real part of the permittivity was then fitted to the derivative of the former expression as

$$-d\varepsilon'(\omega)/d\log\omega = \mathrm{Re}\ \{1/\log e \sum \Delta\varepsilon_j\, a_j\, b_j\, (i\omega\tau_j)^{a_j} \; / \; [1 + (i\omega\tau_j)^{a_j}]^{(b_j+1)}\} , \qquad (2)$$

so that the parameters obtained from the fittings can be directly compared with those obtained from $\varepsilon'(\omega)$ and/or $\varepsilon''(\omega)$ fittings. Although the contribution of Maxwell- Wagner-Sillars-type processes to the real part of the permittivity is not important (see figure S4), this was modeled by an extra term $P/\omega^{-x}$, in the derivative of the real permittivity.

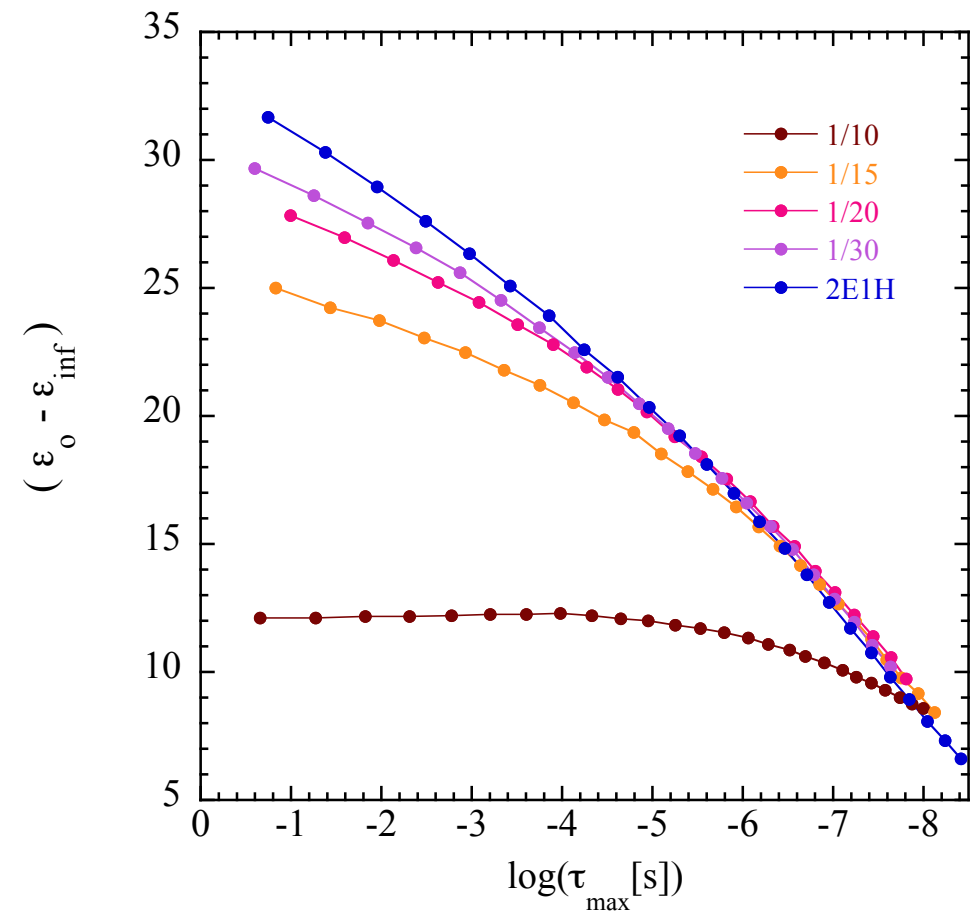

FIG. S7: Evolution of the dielectric strength for different salt concentrations.

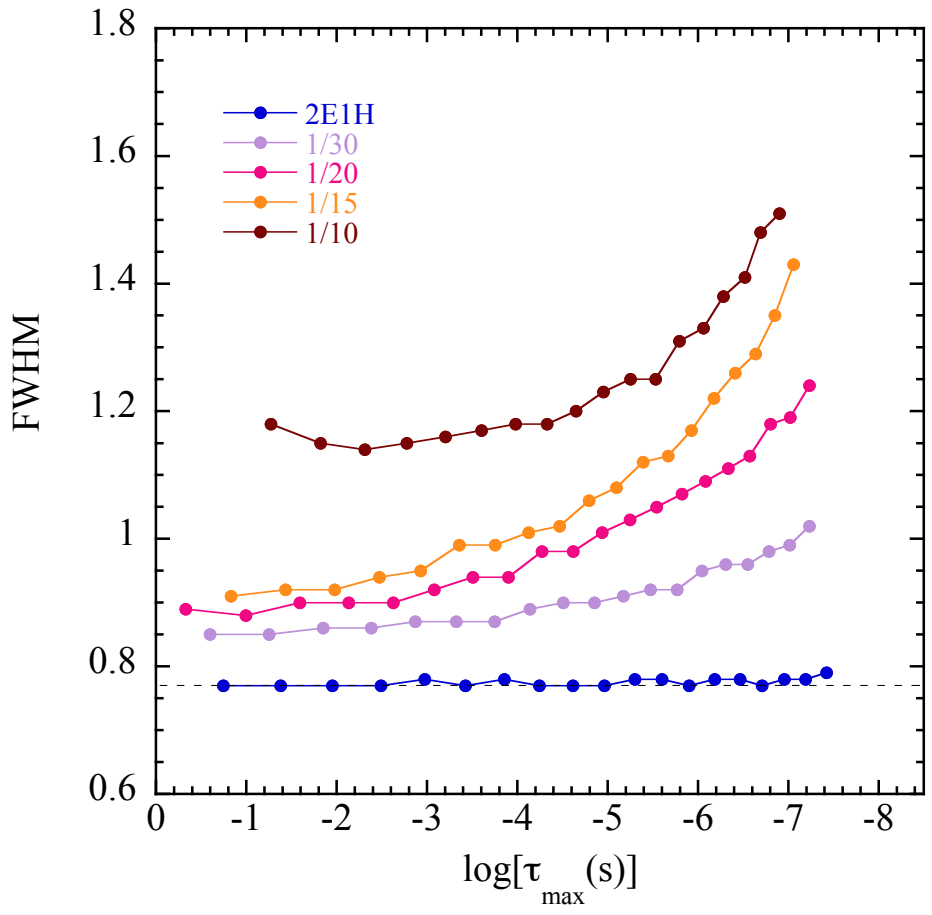

FIG. S8: Full Width at Half Maximum (FWHM) of the derivative of the real part of the permittivity for different salt concentrations. Dashed line indicates FWHM for a Debye (exponential) relaxation.

As at first sight at least dielectric relaxation shows only two relaxations (apart from a faster secondary relaxation not discussed in this paper), a first approach would be the phenomenological description of these two relaxations by the HN expression. Therefore, in first instance two functions were used to describe data: a HN for the slower relaxation (i.e. free *a* and *b*); and a Kolrausch-Williams-Watts-equivalent Havriliak-Negami ($HN_{KWW}$) function [6] for the faster structural relaxation. In the $HN_{KWW}$ function the a and b shape parameters are related to each other as $b = 1 - 0.8121(1 - a)^{0.387}$, and to the equivalent stretching parameter $\beta_{KWW}$ as $\beta_{KWW} \approx (a\ b)^{0.813}$. This 'two HN process approach' describes well the measured data for low temperatures and LiCl concentrations but otherwise shows systematic deviations. Figure S9 shows examples of such fittings for 1/15 sample. As it can be seen, the quality of the fittings gets worse the higher the temperature and the broader the relaxation. At the highest temperatures shown in figure 3 of the manuscript and figure S6 of the present SI the two HN model is not valid any more, an expected result in view of the asymmetry and shape of the experimental relaxation data.

At the light of the conclusions drawn in the paper a quantitative description of data in terms of a model comprising three different processes would be more physically meaning and consistent. Unfortunately, for low concentration samples, the great overlap existing between the slowest relaxation attributed to supramolecular H- bonded structures and the intermediate process prevents the obtention of reliable parameters characterizing each of the relaxations independently. In any case we tested a three process approach as follows. On the one hand, as the low temperature Debye-like relaxation of pure 2E1H is almost an exponential one (equivalent Debye in the frequency domain), the shape of the slowest component was fixed to a Debye function (HN with $a = b = 1$). For the structural relaxation the previously mentioned $HN_{KWW}$ function was used. In view of the great extent of the overlap between different processes we estimated convenient to minimize the number of free parameters with the aim of limiting numerically correct but physically meaningless outcomes. For this reason, in the case of the intermediate process a symmetric Cole-Cole (CC) –HN with $b = 1$– or an asymmetric Cole-Davidson (CD) –HN with $a = 1$– function was used. All in all;

$$\varepsilon^*(\omega) = \varepsilon_\infty + \Delta\varepsilon_{debye}/[1 + (i\omega\tau_{debye})] + \Delta\varepsilon_{int}\ /\ [1 + (i\omega\tau_{int})^{aint}]^{bint} + \Delta\varepsilon_\alpha\ /\ [1 + (i\omega\tau_\alpha)^a]^{1 - 0.8121(1 - a)^{0.387}} \qquad (3)$$

Separate relaxation components resulting form such fits were shown in figure 1 of the manuscript. Temperature evolution of the dielectric strength of the Debye and intermediate processes were also included in figure 4 of the manuscript for the most favorable cases i.e. 1/15 and 1/10 molar ratio samples. For these two samples, 1/15 and 1/10 LiCl/2E1H molar ratio samples, best fits were obtained by a CC function for the intermediate process. In the case of pure MAs, although a CC function also provides good fits at high temperatures where the Debye-like relaxation is broad, numerical minimization sometimes produces physically meaningless outcomes at low temperatures. This is because at low temperatures the relatively narrow character of the Debye-like relaxation does not provide sufficient numerical resolution to obtain a meaningful separate characterization of each component. In these cases, the intermediate process was characterized by a CD function, which somehow limits the characteristic times of this process to a frequency range intermediate between that of the Debye and α processes.

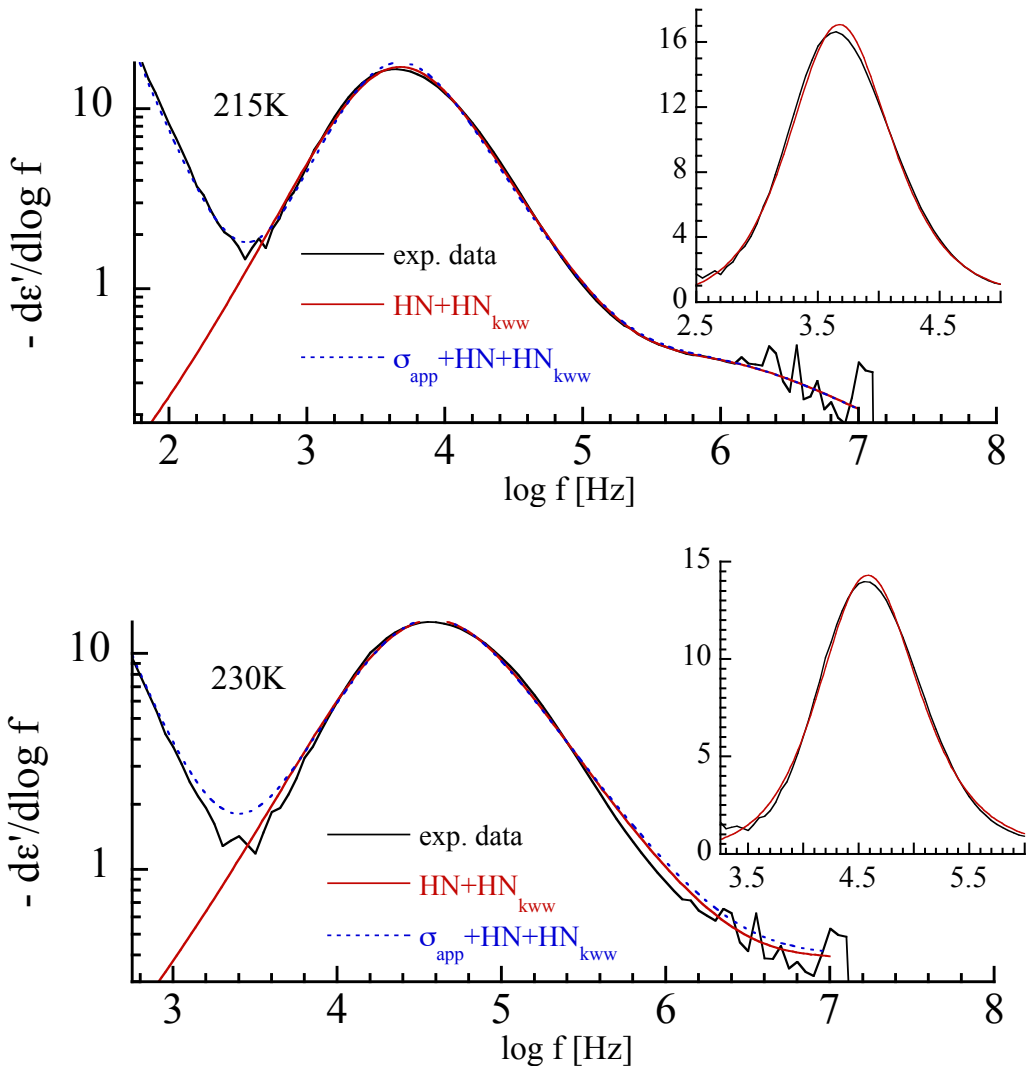

FIG. S9: Fitting of the derivative of the real part of the permittivity of the 1/15 sample to a sum of a HN for the slowest peak and a HNKWW for the faster structural α relaxation. Insets show detail of the fittings at the linear scale evidencing systematic disagreement at the peak maximum.

## References

[1] l. Stordrange, A. A. Christy, O. M. Kvalheim, H. Shen, Y. Liang. Study of the self-association of alcohols by near- infrared spectroscopy and multivariate 2d techniques. *J. Phys.Chem.A* **2002**, 106, 8543-8553.

[2] M. A. Czarnecki and K. Orzechowski. Effect of temper- ature and concentration on self-association of octan-3-ol studied by vibrational spectroscopy and dielectric measurements. *J. Phys.Chem.A* **2003**, 107, 1119-1126.

[3] S. Arrese-Igor, A. Alegría, and J. Colmenero. Dielectric relaxation of 2ethyl-1hexanol around the glass transition by thermally stimulated depolarization currents. *J. Chem. Phys.,* **2015**, 142, 214504.

[4] S. Arrese-Igor, A. Alegría, and J. Colmenero. On the non-exponentiality of the dielectric Debye-like relaxation of monoalcohols. *J. Chem. Phys.,* **2017**, 146, 114502.

[5] F. Kremer and A. Schönhals. *Broadband Dielectric Spectroscopy*. Ed., Springer-Verlag, Berlin, Heidelberg **2003**. ISBN 3-540-43407-0.

[6] F. Alvarez, A. Alegr´ıa, and J. Colmenero. Relationship between the time-domain Kohlrausch-Williams-Watts and frequency-domain Havriliak-Negami relaxation functions. *Phys. Rev. B*, **1991**, 44, 7306.